\documentclass[oldversion]{aa}  
\usepackage{graphicx}
\usepackage{txfonts}
\usepackage{natbib}
\usepackage{epstopdf}
\usepackage{wasysym}
\bibpunct{(}{)}{;}{a}{}{,} % to follow the A&A style

\newcommand{\Brg}{Br$\gamma$ }
\newcommand{\Mdark}{\ensuremath{M_{UDM}}}

\begin{document}
\title{Spectroastrometry of rotating gas disks for the detection of supermassive black holes in galactic nuclei.}
\subtitle{III. CRIRES observations of the Circinus galaxy\thanks{Based on observations made with ESO Telescopes at the Paranal Observatory under programme ID 383.B-0267(A)}}

\author{A. Gnerucci\inst{1}, 
            A. Marconi\inst{1},
           A. Capetti\inst{2},
           D. J. Axon$\dagger$\inst{3,4},
           A. Robinson\inst{3}
}
\offprints{A. Marconi}

\institute{Dipartimento di Fisica e Astronomia, Universit\`a degli Studi di Firenze, Firenze, Italy\\
             \email{gnerucci@arcetri.astro.it, marconi@arcetri.astro.it}
             \and INAF - Osservatorio Astronomico di Torino, Strada Osservatorio 20, 10025 Pino Torinese, Italy\\
             \email{capetti@oato.inaf.it}
             \and Physics Department, Rochester Institute of Technology, 85 Lomb Memorial Drive, Rochester, NY 14623, USA\\
             \email{djasps@rit.edu, axrsps@rit.edu}
             \and School of Mathematical \& Physical Sciences, University of Sussex, Falmer, Brighton, BN2 9BH, UK\\
             $\dagger$ 1951-2012
            }

\date{Received ; accepted}
  
 \abstract{We  present new CRIRES spectroscopic observations of the Br$\gamma$ emission line in the nuclear region of the Circinus galaxy, obtained with the aim of measuring the {black hole (BH)} mass with the spectroastrometric technique. The Circinus galaxy is an ideal benchmark for the spectroastrometric technique given its proximity and secure BH measurement obtained with the observation of its nuclear H$_2$O maser disk.
The kinematical data have been analyzed both with the classical method based on the analysis of the rotation curves and with the new method developed by us and based on spectroastrometry.
The classical method indicates that the gas disk rotates in a gravitational potential resulting from an extended stellar mass distribution and a spatially unresolved dynamical mass of $(1.7\pm 0.2)\times 10^7\,M_\odot$, {concentrated within $r<7$ pc}, {corresponding to the seeing-limited resolution of the observations}. 
The new method is capable of probing the gas rotation at scales which are a factor $\sim 3.5$ smaller than those probed by the rotation curve analysis, highlighting the potential of spectroastrometry. The dynamical mass which is spatially unresolved with the spectroastrometric method is a factor $\sim 2$ smaller, $7.9^{+1.4}_{-1.1}\times 10^6\,M_\odot$ indicating that spectroastrometry has been able to spatially resolve the nuclear mass distribution {down to 2 pc scales}. This unresolved mass is still a factor $\sim 4.5$ larger than the BH mass measurement obtained with the H$_2$O maser emission indicating that, even with spectroastrometry, it has not been possible to resolve the sphere of influence of the BH.
Based on literature data, this spatially unresolved dynamical mass distribution is likely dominated by  warm molecular gas and it has been tentatively identified with the circum-nuclear torus which prevents a direct view of the central BH in Circinus.
This mass distribution, with a size of $\sim 2$pc, is similar in shape to that of the star cluster of the Milky Way suggesting that a molecular torus, forming stars at a high rate, might be the earlier evolutionary stage of the nuclear star clusters which are common in late type spirals.
}

  
   \authorrunning{A. Gnerucci et al.}
     \titlerunning{Spectroastrometry of the nuclear gas disk of the Circinus galaxy.}
     
  \maketitle
  
\section{Introduction}\label{s1}

One of the fundamental yet unresolved problems of modern astrophysics  {relates to} the formation and evolution of the complex structures that characterize the present-day universe such as galaxies and clusters of galaxies. Understanding how galaxies formed and how they become the complex systems we observe today is therefore a current major theoretical and observational effort.

There is now strong evidence for the existence of a connection between supermassive black holes (hereafter BHs), nuclear activity and galaxy evolution. Such evidence, that reveals the so-called co-evolution of black holes and their host galaxies, is provided by the discovery of ``relic'' BHs in the center of most nearby galaxies, and by the tight scaling relations between BH mass ($M_{BH}\sim 10^6-10^{10}M_{\astrosun}$) and the structural parameters of the host spheroids like mass, luminosity and stellar velocity dispersion (e.g. \citealt{kormendy-richstone}, \citealt{Gebhardt:2000a}, \citealt{ferrarese-merrit2000}, \citealt{Marconi:2003}, \citealt{Haring:2004}, \citealt{Ferrarese:2005}, \citealt{Graham:2008}). Moreover, while it has long been widely accepted that Active Galactic Nuclei (AGN) are powered by accretion of matter onto a supermassive BH, it has recently been possible to show that BH growth is mostly due to accretion of matter during AGN activity, and therefore that most galaxies went through a phase of strong nuclear activity (\citealt{Soltan:1982}, \citealt{Yu:2002a}, \citealt{Marconi:2004}). It is believed that the physical mechanism responsible for this coevolution of BHs and their host galaxies is probably feedback by the AGN, i.e. the accreting BH, on the host galaxy (\citealt{Silk:1998}, \citealt{Fabian:1999}, \citealt{Granato:2004}, \citealt{Di-Matteo:2005}, \citealt{Menci:2006}, \citealt{Bower:2006}).
As the scaling relations between BH mass and host galaxy properties represent the clearest sign of co-evolution, it is important to secure these relations by increasing the number, accuracy and mass range of  $M_{BH}$ measurements. 

Supermassive BHs are detected and their masses measured by studying the kinematics of gas or stars in galaxy nuclei.  Currently, there are about $\sim50$ BH mass measurements most of which are in the $\sim10^7-10^9M_{\astrosun}$ range (e.g. \citealt{Sani:2010}). The majority of these measurements have been made with longslit spectroscopy, but the development in recent years of Integral Field Unit (hereafter IFU) spectrographs has resulted in an improvement in accuracy and reliability (see, e.g., \citealt{Davies:2006}, \citealt{Nowak:2007}, \citealt{Nowak:2010},  {\citealt{Krajnovic:2007}, \citealt{Krajnovic:2009}}, \citealt{Cappellari:2009}, \citealt{Neumayer:2010}, \citealt{Rusli:2010}) of the mass determinations.
One crucial issue in BH mass measurements is spatial resolution: a necessary but not sufficient condition is that the spatial resolution must be high enough to make it possible to spatially resolve the central regions of the galaxy, where the gravitational effects of the BH dominate those of the host galaxy (the so-called 'sphere of influence of the black hole'). Even with the advent of Adaptive Optics (AO) assisted observations the best spatial resolutions achievable are $\sim 0.1\arcsec$, which corresponds to $\sim 10$ pc at a distance of $20$ Mpc.

We have developed a new method based on the technique of spectroastrometry which allows gas kinematical BH mass measurements that partly overcome the spatial resolution limitations of the ``classical'' gas (or stellar) kinematical methods. This method has been presented and discussed in detail by Gnerucci et al.~(2010, 2011, hereafter G10, G11), and provides a simple but accurate way to estimate BH masses  using either longslit or IFU spectra.

 {In this paper we apply this spectroastrometric method in an attempt to estimate the mass of the nuclear BH in the Circinus galaxy and to test whether it is possible to resolve the BH sphere of influence.  }
At a distance of 4.2 Mpc (hence $1\arcsec\simeq$ 20 pc), this is one of the closest Seyfert 2 galaxies \citep{Freeman:1977}. Its nuclear activity  is indicated by the observed line ratios \citep{oliva:1994}, by the ionization cone observed in [O III] \citep{marconi:1994a} and [Si VII]   {\citep{Prieto:2005}}, by the narrow coronal lines \citep{oliva:1994,Maiolino:2000}, and by the broad ($FWHM > 3300$ km/s) H$\alpha$ emission  in polarized light \citep{oliva:1998}.
The  discovery of $\mathrm{H_2O}$ maser emission, whose kinematics are indicative of a circularly rotating nuclear  disk, has facilitated one of the most impressive and accurate mass measurements $(1.7\pm0.3)\times10^6 M_{\astrosun}$ \citep{Greenhill:2003,Greenhill:2003a} of a supermassive BH to date. The relatively low BH mass makes the gravitational sphere of influence of the BH very small and spatially unresolved even with AO observations at 8m-class telescopes. These two facts combined make 
this galaxy a very interesting benchmark to test the potential of the spectroastrometric approach. 

Here we present new long-slit spectra of the nuclear region of Circinus obtained with CRIRES at the ESO VLT. This instrument provides a very high spectral resolution that is essential to disentangle the relatively low gas velocities expected due to the relatively low  BH mass estimated by \cite{Greenhill:2003a}.
We perform two parallel analyses and model the CRIRES spectra  using both the classical rotation curves approach and the spectroastrometric method, enabling us to check the quality of the modeling and to put tight constraints on $M_{BH}$ based on a comparison of the independent results.
In Sect.~2 we briefly review the key results of G10 and G11 on the application of spectroastrometry to rotating gas disks to enable the detection of nuclear BHs. In Sect.~3 we present the new CRIRES observations together with details of the data reduction and subsequent analysis. In sect.~4 we describe the CRIRES Circinus gas rotation curves and their modeling. In Sect.~5 we present the matching spectroastrometric analysis and modeling of the CRIRES spectra.  We compare and discuss the results from the spectroastrometric and rotation curves analyses in Sect.~6, with the aim of mutually constraining  the two approaches. Finally, in Sect.~7 we discuss the  results and draw our conclusions.

\section{Spectroastrometric measurements of black holes masses}\label{s3}

The spectroastrometric method (see \citealt{Bailey:1998}) consists in measuring the photocenter of emission lines in different wavelength or velocity channels. 
It has been used by several authors to study pre-main sequence binaries and inflows, outflows or the disk structure of the gas surrounding pre-main sequence stars (\citealt{Takami:2003}; \citealt{Baines:2004}; \citealt{Porter:2004,Porter:2005}, \citealt{Whelan:2005}).

In \citealt{Gnerucci:2010} (G10) we illustrated how the technique of spectroastrometry can be used to measure  black hole masses at the center of galaxies. In that paper we focused on explaining the basis of the spectroastrometric approach and presented an extended and detailed set of simulations showing how this method can be used to probe the principal dynamical parameters of a nuclear gas disk. A key aspect of G10 was the comparison of  this technique with the standard method for gas kinematical studies based on the gas rotation curve. We demonstrated  that the two methods are complementary approaches to the analysis of spectral data (i.e. the former measures mean positions for given spectrum velocity channels while the latter measures mean velocities for given slit position channels).  The principal limit of the rotation curves method resides in the ability to spatially resolve the region where the gravitational potential of the BH dominates with respect to the contribution of the stars. In G10 we demonstrated that spectroastrometry has the ability to provide information on the galaxy gravitational potential at scales significantly smaller than the spatial resolution of the observations  {(potentially reaching $\sim1/10$)}. This fundamental feature can be illustrated by the following simple example: consider two point-like sources located at a distance significantly smaller than the spatial resolution of the observations; these sources will be seen as spatially unresolved with their relative distance not measurable from a conventional image. However, if spectral features, such as absorption or emission lines are present at different wavelengths in the spectra of the two sources, the spatial profiles of the light distribution extracted at these wavelengths will have different centroids, revealing the presence of the two sources. The separation between the sources can be estimated from the difference in the centroid positions at the wavelengths of the different spectral features, even if the separation is much smaller that the spatial resolution. Thus, the ``spectral separation'' of the two sources makes it possible to overcome the spatial resolution limit. 

For clarity, we summarize here the principal features and the main steps of the method presented and discussed extensively in G10.
\begin{itemize}
\item From the longslit spectrum of a continuum-subtracted emission line one constructs the ``spectroastrometric curve'' by measuring the line centroid along the slit for all wavelength channels. The ``spectroastrometric curve'' of the line is given by the position centroids as a function of wavelength.
\item From the simulations presented in G10, we showed that the information about the BH gravitational field is predominantly encoded in the ``high velocity'' (hereafter HV) range of the spectroastrometric curve which comprises the points in the red and blue wings of the line. In the case of Keplerian rotation, the HV part of the line spectrum originates from gas moving at high velocities closer to the BH and this emission is spatially unresolved. For this reason the HV part of the line spectrum is not strongly influenced by the spatial resolution or other instrumental effects like slit losses or by the intrinsic line flux distribution. On the other hand, ``low velocity'' emission (hereafter LV) is usually spatially resolved (i.e.~the gas moving at lower velocities is located farther away from the BH), so the spectroastrometric curve in this regime is not useful for the purpose of measuring the BH mass. The HV range of the curve is identified by measuring the spatial extent of the line emission as function of wavelength and discarding the central (i.e.~LV range) bins where the spatial profile becomes broader than the instrumental spatial resolution.
\item By measuring the spectroastrometric curves of a given line from at least three spectra taken at different slit position angles, one can obtain a ``spectroastrometric map'' of the source on the plane of the sky by geometrically combining the three curves. In the case of integral field spectra, this step is obviously not necessary because the spectroastrometric map is derived directly from the data cubes. In the case of a rotating disk with a radially symmetric line flux distribution the points of the spectroastrometric map should lie on the disk line of nodes. 
However even for IFU data the effect of slit losses or a non-symmetric line flux distribution can perturb the light centroid positions moving them away from the disk line of nodes. As shown in Paper I and discussed above, these effects become negligible for the HV range of the map where the emission is spatially unresolved.  In contrast, the LV points of the map tend to lie away from the line of nodes in a typical ``loop'' shape. The final spectroastrometric map is then obtained by selecting only the HV points.
\item One can then estimate the disk line of nodes by a linear fit to the HV range of the spectroastrometric map, project those points on the estimated disk line of nodes and obtain the disk rotation curve. Finally, one can apply a simple model fitting procedure to obtain the parameters determining the gas rotation curve, and in particular the BH mass.
\end{itemize}

In a subsequent work (Gnerucci et al. 2011, G11), we applied the method to the radio galaxy Centaurus A, using  seeing limited longslit spectra and both seeing limited and AO assisted IFU spectra. We selected this galaxy because it has been extensively studied using the gas kinematical method; the observed kinematics show that the gas is circularly rotating around the BH and the mass and the disk parameters are well constrained. Moreover both longslit and IFU data with different spatial resolutions are available, allowing a direct comparison of the spectroastrometry results to the results obtained from data sets with different characteristics. We demonstrated with real data that the method provides the same mass estimates as the classical gas kinematical method, but allows smaller spatial scales to be probed . 

We briefly summarize the key results of G11 here.
\begin{itemize}
\item
We demonstrated with real data  the capability of spectroastrometry to overcome the spatial resolution limit. The minimum distance from the BH at which we can measure the gas rotation curve is $\sim20mas$, which corresponds to $\sim1/15$ of the spatial resolution for the seeing limited data and $\sim1/6$ for the AO assisted data.
\item The method based on spectroastrometry is in excellent agreement with the classical method based on the rotation curves (the BH mass estimates are consistent within $\pm0.2$~dex and within less than $\pm0.05$~dex when comparing the results obtained from the same dataset). This demonstrates the robustness of the method: using the same dataset we can independently apply the classical and the spectroastrometric methods and obtain consistent results.
\item 
The application of the spectroastrometric method to different types of data (IFU and longslit) give consistent results (within only $\sim0.1$~dex), demonstrating the versatility of our method.
\item
A typical feature of spectroastrometry is its insensitivity to disk inclination. In our modeling, in fact, mass and disk inclination are coupled. Therefore our method can return only a value for $M_{BH}\,sin^2i$; the inclination must be assumed or determined by another method to derive the value of the mass. 
\item
The spectroastrometric method has been applied independently of the classical method. However as discussed in G10, it is clear that the spectroastrometric and ``classical'' rotation curves are complementary and orthogonal descriptions of the position velocity diagram. Therefore, a  development of this method will be its application in combination with the classical method based on rotation curves. This provides  constraints on  the disk inclination and thus removes the mass-inclination degeneracy.
\end{itemize}

The utility of the method is not limited to gas disks rotating around supermassive  {black holes} but also to galaxy disks and indeed we have applied the method to improve virial mass estimates of high-z galaxies observed with IFU spectra (for more details see \citealt{Gnerucci:2011a}a).
%:figure spectrum
  \begin{figure}[!t]
  \centering
  \includegraphics[width=\linewidth]{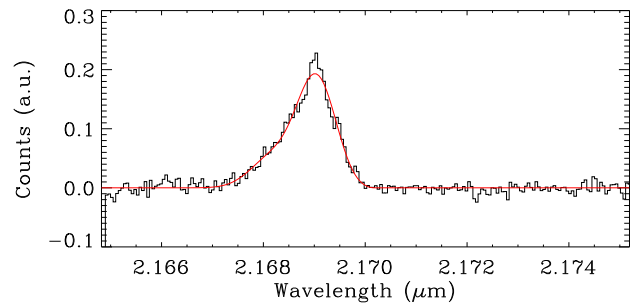}
  \caption{Continuum subtracted detector 3 CRIRES spectrum (slit PA1) extracted at the position of the continuum peak. Solid red line: fit to  \Brg using a Gauss-Hermite function.}
     \label{figspec}
  \end{figure}
%: fig rot
  \begin{figure*}[t!]
  \centering
  \includegraphics[width=0.90\linewidth]{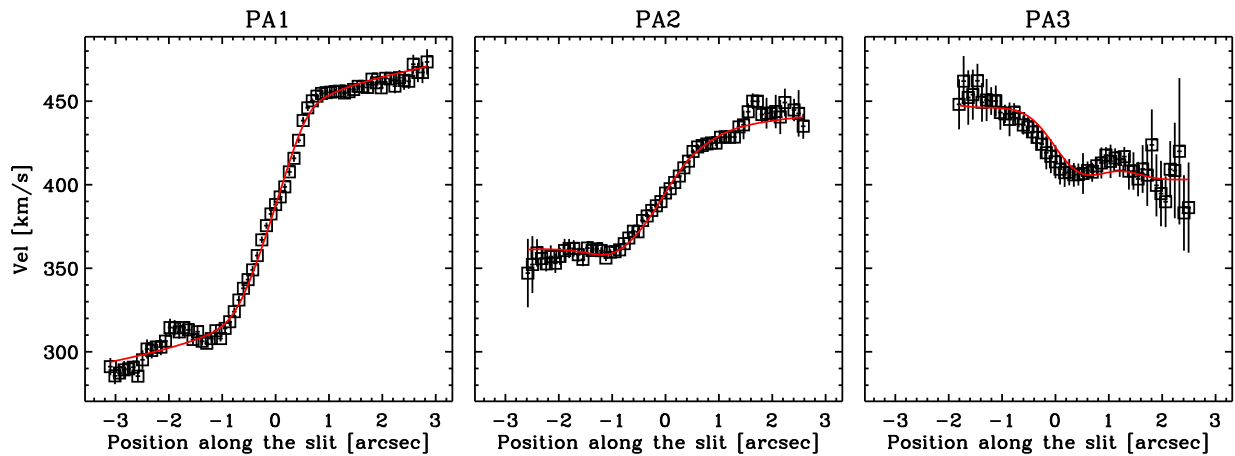}
     \caption{Rotation curves for the CRIRES \Brg spectra. Left panel: slit ``PA1''. Central panel: slit ``PA2''. Right panel: slit ``PA3''. The solid red lines represent the curves expected from the model.}
        \label{figrot}
  \end{figure*}

\section{Observations and data reduction}\label{s41}

Circinus was observed between April and July 2009 (program 383.B-0267A) with CRIRES (Cryogenic high-resolution Infrared Echelle Spectrograph {; \citealt{Kaeufl:2004})} at the ESO VLT with the aid of MACAO (Multi Applications Curvature Adaptive optics) Adaptive Optic system.
The observations consist of three nuclear spectra characterized by a different slit position angle:  {the slits with position angles of $30^{\circ}$, $90^{\circ}$ and $150^{\circ}$ with respect to the north direction are denoted by PA1, PA2  and PA3, respectively. The orientation of PA1 was chosen in order to align the slit with the major axis of the galaxy, and PA2 and PA3 were incrementally rotated by $60^{\circ}$ in order to provide a uniform coverage in position angles with the available slits.} 

 {The Adaptive Optics module MACAO} was used with the unresolved Circinus nucleus as a guide star.
The spectra were obtained with a $0.4\arcsec$ wide slit (slit length $\sim31\arcsec$) and a resolving power of $\lambda/\Delta\lambda=50000$, corresponding to a spectral resolution of $\sim0.4\mathrm{\AA}$ {($\sim 5.5$km/s)} at the observed \Brg wavelength ($2.169\mu m$). The resulting wavelength range is $\sim2.137 - 2.188 \,\mu m$, covered with a mosaic of four InSb arrays providing an effective 4096 x 512 focal plane detector array in the focal plane. The choice of the very high-spectral resolution  is critical for the measurement of the small velocities and velocity dispersions which are expected, given the relatively low BH mass \citep{Greenhill:2003a}. 
The detector pixel scale is $0.086\arcsec/$pixel along the slit axis and $0.102\mathrm{\AA}/$pixel along the dispersion direction.
For each position angle we obtained 6 spectra with an exposure of $180s$, nodding the target along the slit for sky subtraction, resulting in a total on-source exposure of $1080s$.

The data were reduced using the CRIRES reduction pipeline (version 1.8) for the subtraction of dark frames, flat-field correction, sky subtraction, wavelength calibration and for the
combination of the various observing blocks. IRAF tasks\footnote{IRAF is distributed by the National Optical Astronomy Observatories, which are operated by the Association of Universities for Research in Astronomy, Inc., under cooperative agreement with the National Science Foundation.} were then used to  rectify the spectra and to correct for telluric absorption using a suitable standard star.
During data reduction, 
we concentrated only on the third detector that targeted the \Brg line (rest frame wavelength $2.1661\mu m$) observed at $\sim2.169\mu m$.
The spatial resolution of the data was estimated in the final reduced spectra from the extension of the nuclear continuum along the slit  (the galaxy nucleus with an estimated size of $1.9\pm0.6$ pc, corresponding to $0\farcs1\pm0\farcs03$ at 3.5 Mpc,  is in fact unresolved at the spatial resolution of these observations; \citealt{Prieto:2004a}).
We obtain an average value of $\sim 0.7\arcsec$ FWHM for the  spectra at the three PAs indicating that the AO correction did not perform well, probably due to the non point-like nature of the Circinus nucleus in the R band.
The spectra were  continuum subtracted since, as discussed in G10 and G11, the spectroastrometric analysis requires a continuum subtracted spectrum.
Finally, as the dispersion was higher than needed ($\sim1.4$ km/s/pixel at Br$\gamma$), we rebinned the spectra by 5 pixels along the dispersion direction to improve the signal-to-noise ratio.

The final continuum subtracted \Brg spectra have  a typical signal to noise ratio (hereafter S/N) $\sim35$. This value is obtained from the nuclear spectrum extracted from an aperture along the slit equal to the spatial resolution;  the Br$\gamma$ profile is fitted with a Gauss-Hermite model (see below) and the quoted S/N is defined by  the ratio between the  peak model value and the rms of the fit residuals.

With our observational setup  {(10 km\,$s^{-1}$ spectral resolution, 0\farcs7 spatial resolution)} the minimum detectable BH mass is determined by the spatial resolution and the thermal broadening of the ionized gas (i.e. $\sigma \sim 15-22$ km\,$s^{-1}$ for $T_e = (1-2)\times 10^4$ K), and corresponds to
\begin{equation}
M_\mathrm{BH} = \frac{r \sigma^2}{G} = 1.9\times 10^5\, \mathrm{M}_\odot \left(\frac{\sigma}{20\,\mathrm{km/s}}\right)^2  \left(\frac{r}{0\farcs2}\right)
\end{equation}
 {if $r=0\farcs2$ is the spatial scale probed with spectroastrometry. At this stage, this is a purely indicative value.}

  \begin{figure}[t!]
  \centering
  \includegraphics[width=0.75\linewidth]{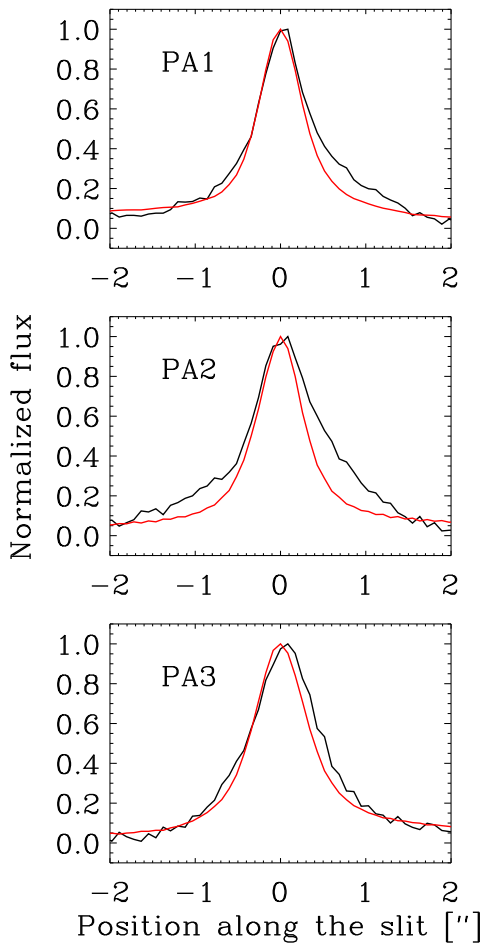}
     \caption{ {Surface brightness profiles  of  the \Brg line along the slits (black lines).  The red lines represent the surface brightness of the underlying continuum.}}
        \label{figflux}
  \end{figure}

%: fig profile
  \begin{figure}[t!]
  \centering
  \includegraphics[width=0.99\linewidth]{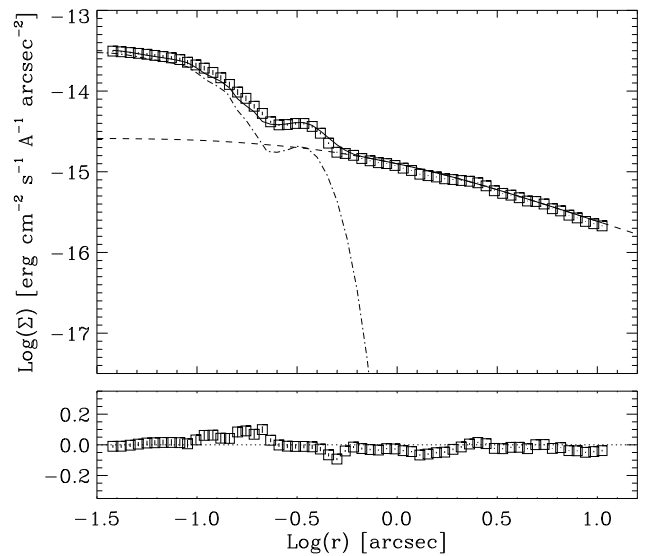}
     \caption{Upper panel. Fit of the NICMOS F222m light profile. The black continuous line represent the best-fit model. The  dot-dashed line represents the profile of the compact component ($0\farcs15$ FWHM size) and the dashed line the extended component. The compact component has been detected and identified with hot dust emission in NACO observations by \cite{Prieto:2004a}. Note that the spatial extension of the compact component is similar to that of the NICMOS PSF and therefore the first Airy ring is still visible. Lower panel: model residuals.}
        \label{figprofile}
  \end{figure}

\section{Analysis of the rotation curves}\label{s42}

We measure the gas rotation curves in the three spectra (PA1, PA2 and PA3) by fitting the \Brg line row by row along the slit direction with a Gauss-Hermite profile \citep{Cappellari:2004} thus obtaining mean velocity, velocity dispersion and line flux at each position along the slit. In Fig.~\ref{figspec} we show a fit example corresponding to the ``PA1'' continuum subtracted spectrum, rebinned by 5 pixels along the dispersion direction and extracted from one pixel along the slit at the position of the continuum peak.
The  rotation curves obtained at the three slit position angles are presented in Fig.~\ref{figrot} and clearly show the typical S-like shapes expected from rotating disks.
 {Fig.~\ref{figflux} shows the \Brg\ surface brightness profiles along the slits (black lines) compared with the corresponding continuum curves (red lines). Since continuum emission is dominated by the unresolved hot dust emission at the spatial resolution of our observations \citep{Prieto:2004a}, figures \ref{figflux} and \ref{figrot}  clearly shows that we  spatially resolve the emission regions and the kinematics of \Brg.}

The rotation curves are first analyzed according to the ``classical'' gas kinematical method. The modeling of the rotation curves is performed following the procedure first described by \cite{Macchetto:1997} and subsequently refined by several authors; 
here we used the modeling code described in detail by \cite{Marconi:2003a,Marconi:2006}. Very briefly, the code computes the rotation curves of the gas assuming that the gas is rotating in circular orbits within a thin disk in the principal plane of the galaxy potential. We neglect any hydrodynamical effect like gas pressure. The gravitational potential is made up of two components: the stellar potential, characterized by its mass-to-light ratio M/L, and a  {spatially unresolved mass concentration} characterized by its total mass \Mdark\ (UDM means ``Unresolved  {Dynamical} Mass'').  {This spatially unresolved dynamical mass concentration is dark in the sense that it is not accounted for by the continuum emission used to trace the stellar mass distribution. It could represent the putative supermassive black hole but also any spatially unresolved stellar or gaseous component or a combination of them}.
In computing the rotation curves we take into account the finite spatial resolution of the observations, the intrinsic surface brightness distribution of the emission lines (hereafter ISBD) and we integrate over the slit and pixel area. The free parameters characterizing the best fitting model are found by standard $\chi^2$ minimization.

We first determined the stellar gravitational potential from a K-band surface brightness profile of the Circinus Galaxy which was obtained from a NICMOS/CAMERA2 F222M mosaic kindly provided by Roberto Maiolino (for details on observations and data reduction see \citealt{Maiolino:2000}).
The inversion procedure to derive the stellar density distribution from the surface brightness is not unique if the gravitational potential does not have spherical symmetry. Assuming that the gravitational potential is an oblate spheroid, the inversion requires knowledge of the potential axial ratio q, and the inclination of its principal plane with respect to the line of sight, $i$. 
These two quantities are related by the observed isophote ellipticity, and for the Circinus galaxy we solved this ambiguity by imposing an axial ratio of $\sim0.1$ that corresponds to an almost disk-like spheroid; Circinus is in fact an SA(s)b spiral galaxy and, moreover, the stellar contribution to the gravitational potential is maximized by imposing a disk-like geometry. 
Following \cite{van-der-Marel:1998}, we then converted the oblate spheroid density distribution model to an observed surface brightness distribution in the plane of the sky by integrating along the line of sight, convolving with the Point Spread Function (PSF) of the telescope+instrument system and averaging over the detector pixel size. Then, the derived model light profile can be directly compared with the observed one. A detailed description of the relevant formulae and the inversion and fit procedure is presented in \cite{Marconi:2003a}.
The  results of the analysis of the NICMOS F222M profile are presented in Fig.~ \ref{figprofile}. In the upper panel the black solid line represents the model surface brightness profile convolved for the instrumental effects. This model profile is the combination of an extended and a compact components, represented by dashed and dot-dashed lines, respectively.
The extended component represents stellar emission and is parameterized with the oblate spheroid described above.  The compact component  is characterized by spherical symmetry  and by a sharp boundary, resulting in  a FWHM of $0\farcs15$. This component is consistent with the compact K band source of size $\sim 0\farcs2$ detected with NACO by \cite{Prieto:2004a} and identified with hot dust emission from the AGN torus on the basis of the steep near-IR colors. This identification has been further confirmed by  \cite{Muller-Sanchez:2006} who were able to deconvolve the stellar and dust continua, confirming that the compact K band emission is dominated by hot dust. Therefore we have not considered it for the stellar mass profile used in the analysis of the rotation curves.

The best fitting rotating disk model is represented in Fig.~ \ref{figrot} by the red solid line and, overall, it nicely reproduces the observed rotation curves.
One possible cause of concern is that the rotation curve at PA3 shows a $\sim50km/s$  velocity shift with respect to the rotating disk model and, indeed, the red line in the right panel of Fig.~ \ref{figrot} has been offset by the same amount to match the observed points. PA3, with a position angle of  $150^\circ$ is almost perpendicular to the line of nodes, which approximately aligns with the PA1 slit (P.A. $30^\circ$).
This shift might then be accounted for by outflowing motions since the PA3 slit is lying within the ionization cone illuminated by the central AGN.
Such outflows are commonly observed in Seyfert galaxies and are likely driven by AGN radiation pressure or winds (e.g. \citealt{muller-sanchez:2011} and references therein). However, since the rotation curves from the PA1 and PA2 slits are both well reproduced by a rotating disk model and since, after allowing for the velocity shift, the PA3 slit rotation curve is also well reproduced by the same model, we believe that the outflowing motions detected in PA3 do not affect the final mass estimate, similarly to what has been found in Cygnus A \citep{Tadhunter:2003}.

\begin{figure}[t!]
  \centering
  \includegraphics[width=0.90\linewidth]{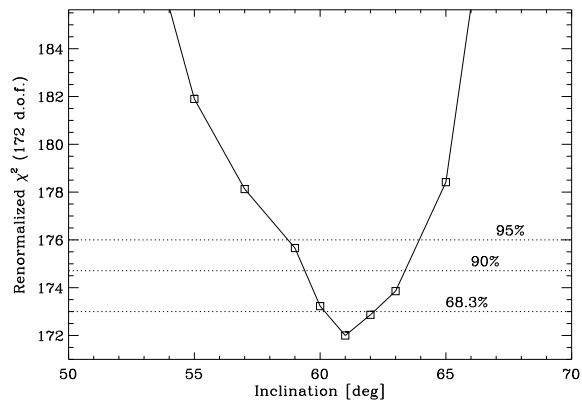}
     \caption{$\chi^2$ vs. $i$ curve for the determination of best value and confidence interval of the disk inclination. The dotted lines represents the $1$, $2$ and $3\sigma$ confidence levels.}
        \label{figinc}
  \end{figure}

Since the disk inclination is coupled with the unresolved dynamical mass (i.e.~ \Mdark\ and stellar M/L ratio, see, e.g., \citealt{Marconi:2006}), we choose to keep it as a fixed parameter in the modeling to avoid convergence problems with the $\chi^2$ minimization. Its best value and confidence interval is then identified by using a grid of fixed $i$ values and performing the fit for each of them. In this way we construct the $\chi^2$ vs. $i$ curve shown in  Fig.~\ref{figinc} which allows us to identify both the best i value and the confidence intervals (for details see \citealt{Avni:1976}). The best inclination value and confidence interval (at $1\sigma$ confidence level) is $(61\pm1)^{\circ}$. This value is in good agreement with the value of $65^{\circ}$ recovered by \cite{Freeman:1977} from large-scale optical photometry and with the value of $55^{\circ}$ obtained by \cite{Hicks:2009} for the inner  $\mathrm{H_2}$ gas disk (radius $\lesssim9pc$).

%: fig inc

The best-fit model parameters for the \Brg rotation curves are reported in Table.~\ref{tab1}, while in Fig.~\ref{figrot} we show the \Brg\ rotation curves and the relative best-fit model.
We estimate the uncertainties on the best-fit model parameters using the bootstrap method (\citealt{efron:1994}). The dataset consists of 181 elements in total from the rotation curves of the three slits, each characterized by spatial position, flux, velocity and velocity dispersion. We randomly extracted from this dataset a subsample with the same number of elements. Due to the random extraction, each subsample will have some elements replicated a few times and some elements that are entirely missing. We then performed the fit on these randomly extracted 100 subsamples and  estimated errors on parameters by taking the standard deviation of the best fit values, usually normally distributed.

\begin{figure}[t!]
  \centering
  \includegraphics[width=0.90\linewidth]{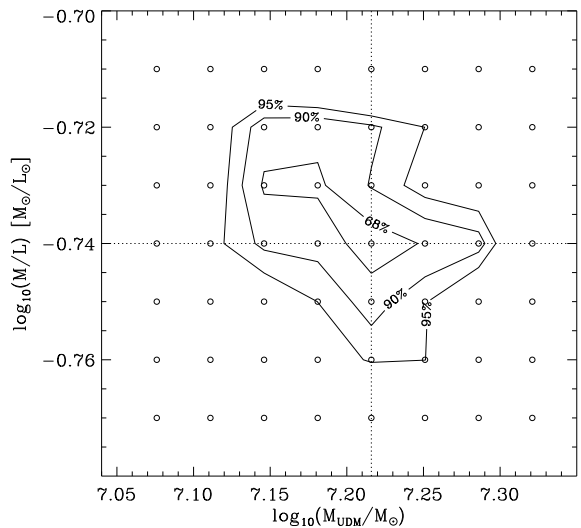}
     \caption{ Confidence regions in the \Mdark-$M/L$ plane for the rotation curve model fits. The iso-contours indicate the $95\%$, $90\%$ and $68\%$ (the inner one) confidence levels. The dotted line marks the \Mdark-$M/L$ values corresponding to the minimum $\chi^2$ fit. The open circles indicate the grid of values used in the \Mdark-$M/L$ plane.}
        \label{figmbhml}
  \end{figure}

Since  \Mdark\ and $M/L$ are correlated, we have also evaluated the confidence intervals for the coupled distribution of these two model parameters. To  {accomplish} this we constructed a grid of values in the $\Mdark-M/L$ plane and performed the fit for each of these points by varying all the other parameters. In this way we created a $\chi^2$ surface which allows us to identify both the best values and the confidence regions in the $\Mdark-M/L$ plane(for details see \citealt{Avni:1976}). In Fig.~\ref{figmbhml} we show the resulting contours of the confidence regions and each contour is labeled with the corresponding confidence level.
 As shown in the figure, the confidence regions are consistent with the statistical uncertainties obtained with the bootstrap method for these parameters (cf. Table \ref{tab1}) and the \Mdark\ and $M/L$  parameters show the expected anti-correlation.

%: fig mbhml

\section{Spectroastrometric analysis}\label{s43}

\subsection{Analysis of the spectra}\label{s431}

As discussed in G10 and G11 to reduce the effects of noise in measuring the spectroastrometric curves, we constructed  a ``synthetic'' position velocity diagram (hereafter PVD) of the \Brg\ line for each longslit spectrum.
We fitted the \Brg line with a Gauss-Hermite function at all positions along each the PA1, PA2 and PA3 slits, obtaining the relative model parameters (flux, mean velocity, velocity dispersion and Hermite parameters $h_3$ and $h_4$). We then use the fitted profiles of \Brg\ to reconstruct noise-free synthetic PVD's for each slit. 
  
\begin{figure}[!h]
  \centering
  \includegraphics[width=\linewidth, trim=10 0 0 0]{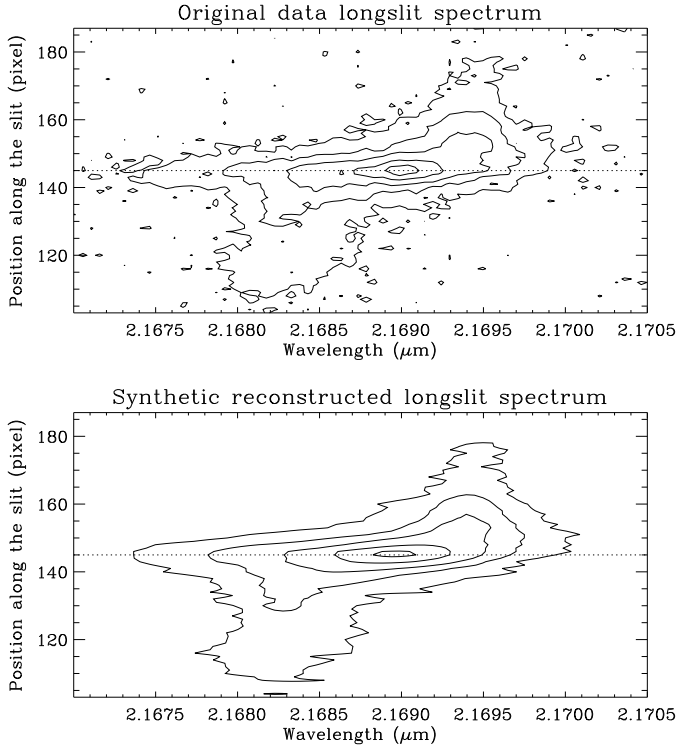}
  \caption{Position velocity diagram of the continuum subtracted \Brg spectrum at PA1. Upper panel: observed. Bottom panel: ``synthetic'' reconstruction. The horizontal dotted line overplotted on each panel represents the continuum peak position. The isophotes denote the same values in both panels.}
        \label{figsynt}
  \end{figure}

{  The PVD for the original \Brg longslit spectrum at PA1 {is compared} with its ``synthetic'' version in Fig. \ref{figsynt}.} It can be seen that the ``synthetic'' PVD is noise-free because each row represents the fitted parametric profile. However, one has to take into account the errors on the fitted parameters in order to estimate the errors on the synthetic pixel counts. Therefore, for each spectral fit, we simulated $1000$ synthetic spectra from 1000 realizations of the set of the five profile parameters distributed following a pentavariate distribution with the fit correlation matrix. The flux and error of each pixel is then estimated from the mean and standard deviation of the 1000 realizations.  We note that for each flux profile along the slit, the errors on fluxes are uncorrelated because they originate from independent fits to the line profiles.

%
%:figure synt
%
  %: fig 4
  \begin{figure*}[!ht]
  \centering
  \includegraphics[width=\linewidth]{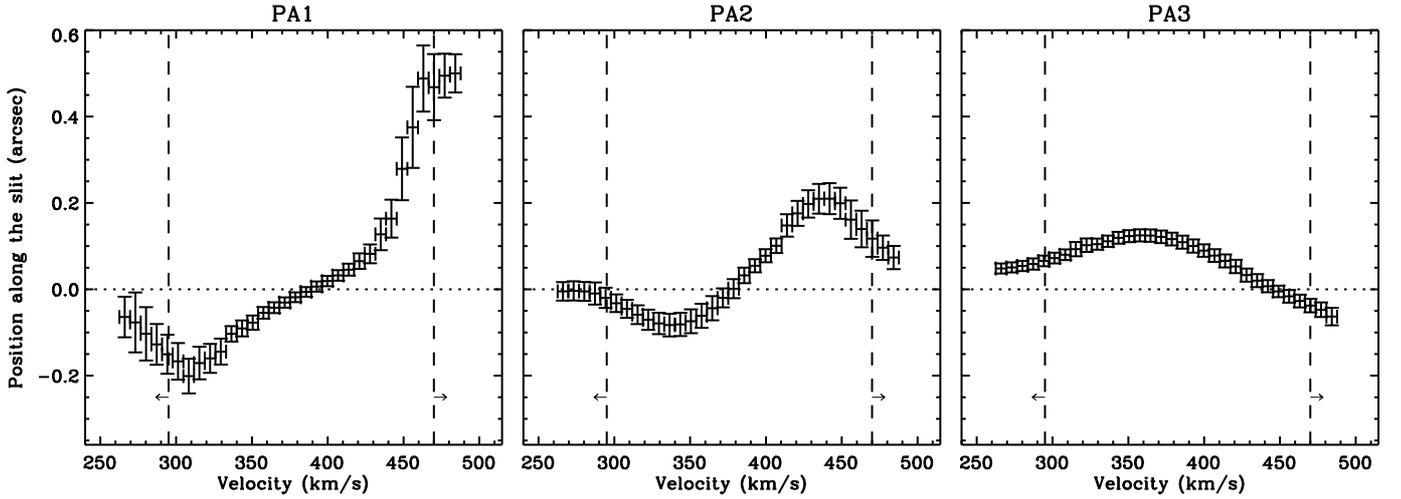}
     \caption{Spectroastrometric curves of the \Brg line at the three slit position angles. Left panel: PA1. Central panel: PA2. Right panel: PA3. The dashed vertical lines on each panel represent the limits of the ``high velocity'' range. Wavelengths have been converted into velocity using as reference (zero velocity) the rest frame \Brg wavelength ($2.1661\mu m$).}
        \label{figspecast}
  \end{figure*}

  From the ``synthetic'' \Brg spectra for the three slits ($PA1$, $PA2$ and $PA3$), obtained following the method outlined above, one can derive the spectroastrometric curves which are shown in Fig.~\ref{figspecast}. 
  
From Fig. \ref{figspecast} we can see that the position of the light centroids in the high velocity (HV) range tend to approach the $0$ position (i.e. the position of the continuum peak).  As discussed in G10 and G11, this is the expected behavior of  spectroastrometric curves in the presence of an unresolved central mass (i.e.~the BH). 
 {This behavior is less evident on the red side because of the small number of high velocity points}.
Errors on photocenter positions range from $\sim 0.02\arcsec$ to $\sim0.1\arcsec$, that is from $\sim 1/35$ to $\sim 1/5$ of the spatial resolution of the data; this is the accuracy with which we can measure centroid positions.

\subsection{The spectroastrometric map of the source}\label{s432}

\begin{figure}[!ht]
  \centering
  \includegraphics[width=\linewidth]{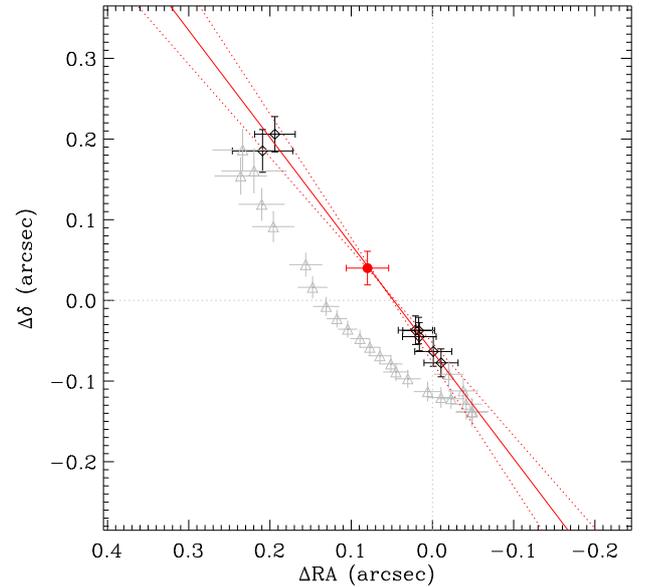}
  \caption{2D spectroastrometric map for the \Brg line. The red point marks the inferred  position of the point-like mass. The red solid line represents the line of nodes of the disk obtained from a linear fit to the HV points, as described in the text. The dotted red lines represent the $1\sigma$ uncertainties on the line of nodes position angle. The origin of the map is the position on the sky plane of the PA1 slit center (which corresponds to the continuum peak position).}
    \label{fig2dm}
  \end{figure}

We  obtained three spectroastrometric curves from \Brg, one for each PA of the slit.  Each spectroastrometric curve provides the
photocenter position along one slit, i.e. the position of the photocenter projected along the axis defined by the slit direction. Thus, combining the spectroastrometric curves obtained at different PA, we can  obtain the map of photocenter positions on the plane of the sky for each velocity bin. In principle, the spectroastrometric curves from two non-parallel slits should suffice but we can use the redundant information from the three slits to recover the 2D sky map as described in Section 4.1 of G10. We have chosen a reference frame in the plane of the sky centered on the center of the PA1 slit (which corresponds to the position of the continuum peak along the slit) with the X axis along the North direction. For a given velocity bin we then determined the position of the light centroid on the sky plane, producing the 2D spectroastrometric map shown in Fig.~\ref{fig2dm}.  

%: fig 2dmap

As discussed in G10 and G11 the coordinates on the plane of the sky of the center of the PA2 and PA3 slits must be considered as free parameters. These unknowns are estimated simultaneously with the position of the photocenter following a $\chi^2$ minimization procedure. The final spectroastrometric map on the plane of the sky is that given by the best
fitting set of slit centers.  The error bars on the points represent the uncertainties resulting from the fit. The black points correspond to the ``high velocity'' points  {(i.e. $v\lesssim295$ km$/s$ and $v\gtrsim475$ km$/s$)} which were actually used to determine the location of the slit centers. 
The HV parts of the line profile represent the tails of the line-of-sight velocity distribution and could not be selected by identifying spatially unresolved emission along the slit, i.e.\ the criterion adopted in G11, since emission along the slit direction is  spatially resolved.
We therefore extracted a spectrum of the Br$\gamma$ line by averaging the CRIRES spectrum over an aperture equal to the spatial resolution and centered on the continuum peak and we  selected the velocity bins characterized by a flux lower than $0.3$ times the line peak flux. This criterion to identify the HV range is different from the one used in G11 for the Centaurus A spectra. However, it produces the same HV range for Centaurus A and is therefore equivalent.
The spatial extension of the emission in the HV region most probably indicates that the high velocity spectroastrometric measurement probes the gravitational potential of an extended mass distribution  {rather} than that of the BH. This issue will be discussed below, where we will show that, indeed, the spectroastrometric analysis indicates the presence of an extended mass distribution. The emission of the gas moving at higher velocities does not originate from the inner region of the disk rotating around the BH but feels the gravitational potential of  an extended mass component.

  The two-dimensional spectroastrometric map just described (and shown in Fig.~\ref{fig2dm}) can now be used to estimate the geometrical parameters of the nuclear gas disk. If the gas kinematics are dominated by rotation around a point-like mass (\Mdark), the position of the light centroid at the high velocities should lie on a straight line (which identifies the direction of the disk line of nodes) and should approach, at increasing velocities, the position of the rotation center. This consideration allows us to make a first estimate of the position on the plane of the sky of the rotation center as the average position of the HV points in the 2d spectroastrometric map\footnote{In practice, we calculate this position by first taking the averages of the coordinates of the points in the blue and red HV ranges and then taking the average coordinates of the ``blue'' and ``red'' positions.}  as well as the position angle of the line
of nodes ($\theta_{LON}$), obtained by fitting a straight line to the high velocity points (see Fig. \ref{fig2dm}). A more accurate estimate of these parameters, derived from the model fitting, will be described in the next section (see Table \ref{tab1}). The uncertainty on center position is $\sim0.04\arcsec$, i.e.~$\sim1/18$ of the spatial resolution of the data.

\subsection{Estimate of the unresolved mass from the spectroastrometric map}\label{s433}

We recover the value of the point-like mass from the spectroastrometric map, following the method presented and discussed in detail in G10 and G11.

Briefly, under the assumption that the gas lies in a thin disk configuration inclined by $i$ with respect to the plane of the sky ($i=0$ face-on) and the disk line of nodes has a position angle $\theta_{LON}$, the line of sight component  of the circular velocity of a gas particle with distance $r$ from the rotation center (hereafter $V_{ch}$ for ``channel velocity'') is given by:

\begin{equation}
V_{ch}=\pm \sqrt{\frac{G[\Mdark+M/L\cdot L(r)]}{r}} \sin(i)+V_{sys}
\label{1}
\end{equation}

where  $L(r)$ is the radial luminosity density distribution in the galactic
nucleus, $M/L$ is the mass to light ratio of the stars and we also added the systemic velocity of the galaxy $V_{sys}$.

First we recover the disk line of nodes by fitting a straight line on the 2d map. Then we project the position of the 2d map points  $(x_{ch}, y_{ch})$ on the line of nodes, calculating their coordinate with respect to this reference axis  ($S_{ch}$) and then their distance $r$ from the rotation center used in Eq. \ref{1} (i.e. $r=k|S_{ch}-S_0|$ where $S_0$ is the coordinate along the line of nodes of the disk and $k$ is a scale factor to transform angular to linear distances\footnote{For consistency with previous works we assume a distance to Circinus of 4.2 Mpc. At this distance $1\arcsec$ corresponds to $\sim20.4pc$.}; see G10 for details).

The unknown parameters of this model are found by minimizing the quantity

\begin{equation}
\chi^2=\sum_{ch}{\left[\frac{V_{ch}-{{V}_{ch}}^{model}}{\Delta(S_{ch}; par)}\right]^2}
\label{4}
\end{equation}

where $\Delta(S_{ch}; par)$ is the uncertainty  of the numerator. As previously discussed, we restrict the fit (i.e. the sum over the velocity channels) to the HV range of the map. The channel velocity $V_{chan}$ has no associated uncertainty since it is not a measured quantity but the central value of the velocity bin. Finally we add a constant error ($\Delta_{sys}$) in quadrature to the quantity $\Delta(S_{ch}; par)$ in order to obtain a reduced $\chi^2$ close to $1$ (see sect.~5 of G10 for a detailed explanation of this choice).
Finally, using the best fit values of the model parameters we can compute the $(x,y)$ position of the rotation center in the sky plane.

We verified that the value of \Mdark\  is  affected little by the actual values of the position angle of the line of nodes and of $S_0$, the position of the rotation center along the line of nodes. Indeed, repeating the fit with $\theta_{LON}$ and $S_0$ values randomly varied within the uncertainties, \Mdark\ changes by $\sim0.06$~dex, consistently with the typical $1\sigma$ statistical uncertainties of the fit (see Table \ref{tab1}).

As discussed previously and in G10 and G11, disk inclination $i$ and  {spatially unresolved mass concentration} are coupled and this fitting method can only measure $\Mdark\,\sin^2i$. Therefore,  we assumed a value for the inclination to obtain a mass value. In the following, {we will consider a  disk inclination ${i=61^{\circ}}$,}  as obtained in the rotation curve analysis.
Summarizing, the free parameters of our fit are:
\[
  \begin{array}{lp{0.8\linewidth}}
     \Mdark     & spatially unresolved mass;     \\
     M/L         & mass to light ratio of the nuclear stars;\\
     S_0      & position of the rotation center along the line of nodes ;\\
     V_{sys}      & systemic velocity of the galaxy;\\
  \end{array}
\]

\begin{figure*}[ht!]
  \centering
  \includegraphics[width=0.48\linewidth]{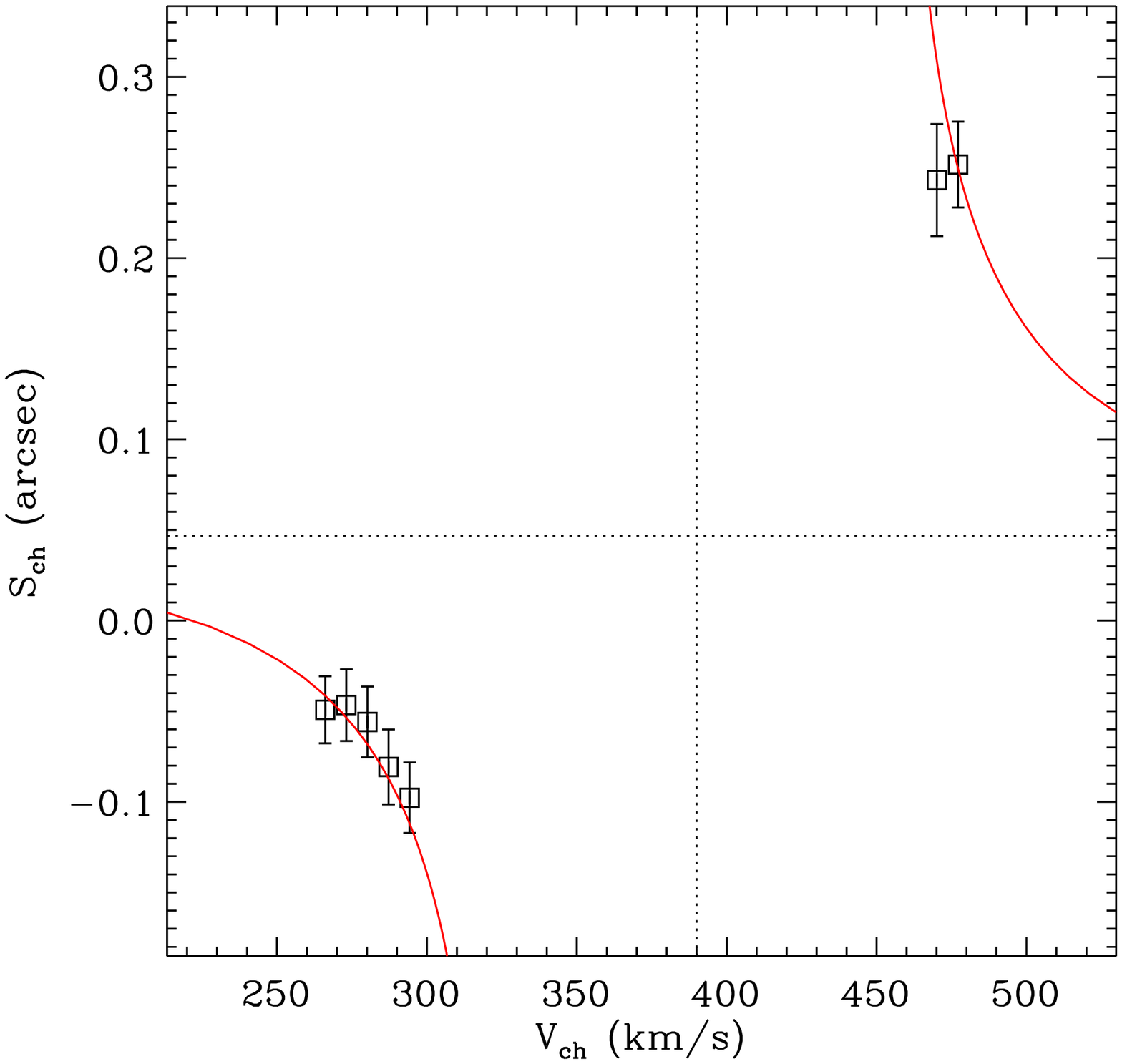}
  \includegraphics[width=0.48\linewidth]{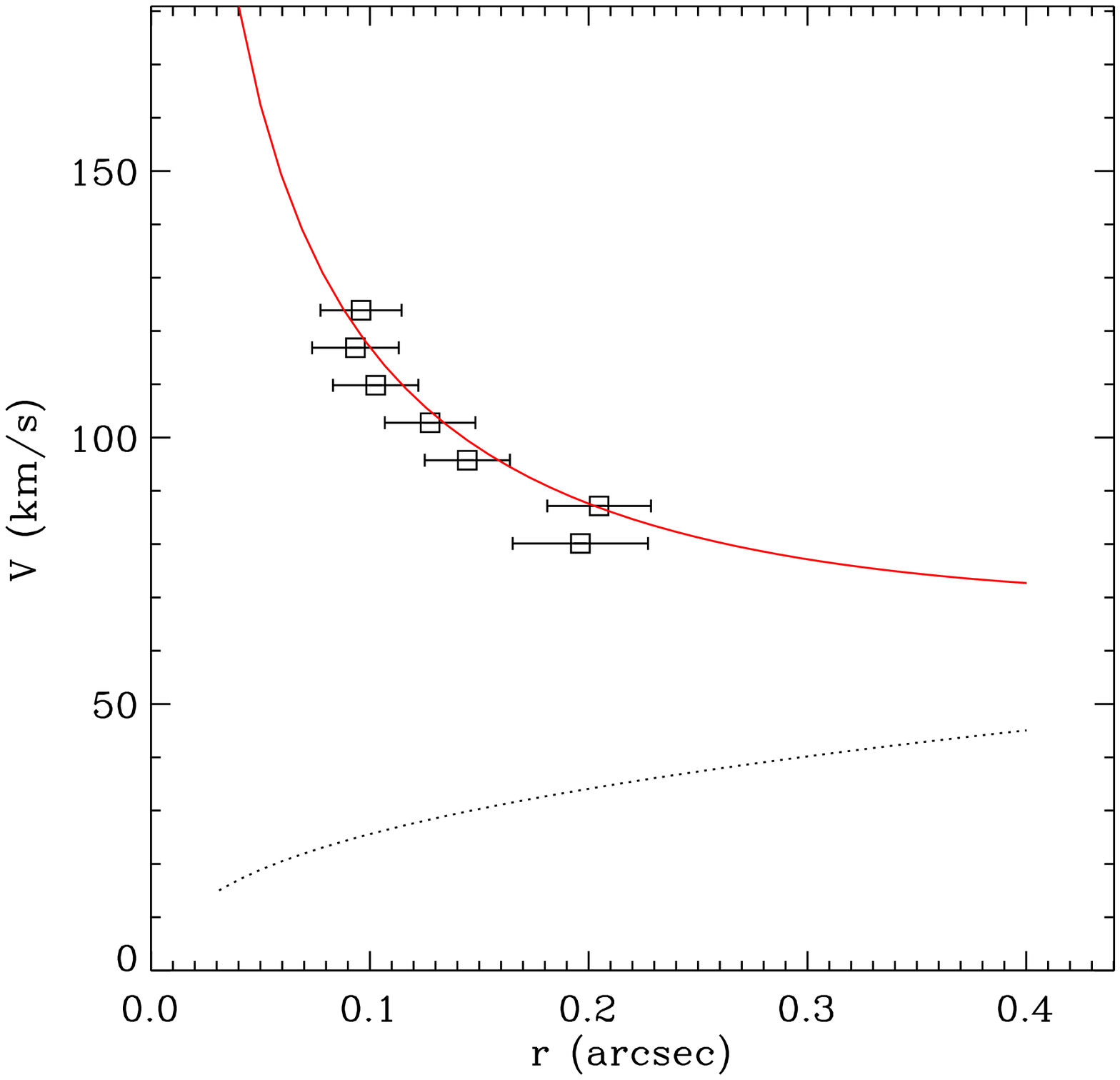}
     \caption{Results of the fit to the spectroastrometric \Brg data. The left panels show the line of nodes projected rotation curve $S_{ch}$ vs. $V_{ch}$. The right panels show the $r=|S_{ch}-S_0|$ vs. $V=|V_{ch}-V_{sys}|$ rotation curve. The solid red lines represent the curves expected from the model. The dotted line on right panel represents the rotation curve due to the star component of the gravitational potential.}
        \label{figfit}
  \end{figure*}

The fit results are tabulated in Table \ref{tab1} and presented graphically in Fig. \ref{figfit}, where we plot the $r=|S_{ch}-S_0|$ vs $V=|V_{ch}-V_{sys}|$ rotation curve and the line of nodes projected rotation curve (e.g.  $S_{ch}$ vs $V_{ch}$). The solid red lines represent the curves expected from the model  ($r$ vs $|{{V}_{ch}}^{model}-V_{sys}|$ and $S_{ch}$ vs ${{V}_{ch}}^{model}$ respectively).

%: fig fit

The first important result from this analysis is that the minimum distance from the rotation center at which there is a velocity estimate 
is ${\sim0.1\arcsec}$ corresponding to $\sim2.3$ pc (see right panel of Fig.~\ref{figfit}), while with the standard rotation curve method the minimum distance from the center which can 
be observed is of the order of half the spatial resolution ($\sim0.35\arcsec$). This clearly shows how spectroastrometry can overcome the spatial resolution 
limit.

 {It is not possible to obtain measurements at  spatial scales smaller than $\sim 1/3$ of those probed by the rotation curve analysis because of the limited signal-to-noise ratio of the spectra; this prevents us from measuring the photocenters of the line at higher velocities, further extending the curves in Fig.~\ref{figspecast} to the red and to the blue sides.}
  
%__________________________________________________ Two column table
%: table 1
\begin{table*}
   \caption[!h]{Fit Results.}
   \label{tab1}
   \centering
   \begin{tabular}{l c c c c c}
     \hline
     \noalign{\smallskip}
     Parameter & \multicolumn{4}{c}{Best fit value$\pm$error} \\
     \noalign{\smallskip}
     \hline
     \noalign{\smallskip}
       & \multicolumn{2}{c}{\Brg Rotation Curves Modeling $^{(1)}$} & \multicolumn{2}{c}{\Brg Spectroastrometric Modeling} \\
     \noalign{\smallskip}
     & Fit-R 0 & Fit-R 1 & Fit-S 0 & Fit-S 1 \\
     \hline
     \noalign{\smallskip}
     $\theta_{LON}\ \ \ [^{\circ}]$      & $215.0\pm0.5$  & $217\ ^{(2)}$ & $217[\pm4]{^{(3)}}$  & $217[\pm4]{^{(3)}}$ \\
     \noalign{\smallskip}
     $x_c\ \ \ [\arcsec]$        & $-0.013\pm0.02$ & $0.04\ ^{(2)}$ & $0.04\pm 0.04$ & $0.04\pm 0.03$ \\
     \noalign{\smallskip}
     $y_c\ \ \ [\arcsec]$        & $0.067\pm0.02 $ & $0.08\ ^{(2)}$ & $0.08 \pm 0.03{^{(4)}}$ & $0.08 \pm 0.02{^{(4)}}$ \\
     \noalign{\smallskip}
     \noalign{\smallskip}
     \noalign{\smallskip}
     \noalign{\smallskip}
     $log_{10}(\Mdark/M_{\astrosun})$ & $7.22\pm0.05$ & $7.19$ & $6.93\pm 0.07$ & $6.88\pm 0.07$ \\
     \noalign{\smallskip}
     $log_{10}(M/L)$                                 & $-0.74\pm0.01$ & $-0.737$ & $-11.2\pm 0.0\ ^{(5)}$ & $-0.74\ ^{(2)}$ \\
     \noalign{\smallskip}
     $i\ \ \ [^{\circ}]$                                & ${61[\pm 1]^{(3)}}$ & ${61[\pm 1]^{(3)}}$ & ${61[\pm 1]^{(3)}}$ & ${61[\pm 1]^{(3)}}$\\
     \noalign{\smallskip}
     $V_{sys}\ \ \ [km/s]$                              & $389.3\pm0.8$ & $393.6$ & $391\pm16$ & $390\pm9$ \\
     \noalign{\smallskip}
     %$\Delta_{sys}\ \ \ [km/s]$                             &$-^{(2)}$  & $-^{(2)}$ & $0$ & $0$ \\
     \noalign{\smallskip}
     \noalign{\smallskip}
     \noalign{\smallskip}
     \noalign{\smallskip}
     $\chi^2_{red} \ \ \ (\chi^2/D.O.F.)$ & $1.80\ \ (309/172)$ & $3.69\ \ (645.4/175)$ & $0.25\ \ (0.76/3)$ & $0.3\ \ (1.4/4)$ \\
     \noalign{\smallskip}
     \hline
   \end{tabular} 
\tablefoot{\tablefoottext{1}{The best-fit parameter confidence intervals are computed only for the interesting parameters as explained in the text (see also \citealt{Avni:1976}). }
%\tablefoottext{2}{Parameter not present in this model. }
\tablefoottext{2}{Parameter held fixed. } {\tablefoottext{3}{Parameter held fixed. The adopted value (with corresponding errors) was estimated independently from the fit presented in the table.}}  {\tablefoottext{4}{Derived from $x_c$ using the adopted value of $\theta_{LON}$. }} \tablefoottext{5}{Parameter not constrained from the fit. }}
\end{table*}

\section{Comparison of the rotation curves and spectroastrometric analyses}

Here we compare the results of the two independent analyses of the CRIRES spectra (i.e.~ the fits labeled as ``Fit-R 0'' and ``Fit-S 0'' in Table \ref{tab1}) and discuss the combined constraints from the two methods. Eventually, we will re-analyze the data taking into account these combined constraints.

%----------constraining della M/L da rotcurve
As discussed in G10 and G11, the modeling of the spectroastrometric data is not able to constrain the mass to light ratio $M/L$ of the nuclear star distribution because it models the gas rotation curves at small distances from the rotation center where the contribution of the stellar mass to the gravitational potential is negligible. In fact performing a first model fit with $M/L$ free to vary (``Fit-S 0'' in Table \ref{tab1}) we obtain a value of $log_{10}(M/L)\simeq-11$. Therefore we used the results of the rotation curve modeling to constrain this parameter. In the ``Fit-S 1'' presented in Fig.~\ref{figfit} and Table \ref{tab1} we fixed M/L to the value obtained from the rotation curves analysis.
%----------confronto dei parametri geometrici
Another important check concerns the geometrical parameters of the modeled gas disk (i.e.~the position angle of the line of nodes  $\theta_{LON}$ and the disk center, or equally, the center position on the plane of the sky $(x_c,y_c)$).  The spectroastrometric model fitting gives a value of $217^{\circ}$ for $\theta_{LON}$. This is consistent with the rotation curve modeling, which gives a best-fit value of $215^{\circ}$. On the other hand, the rotation curve modeling gives a rotation center that is shifted south by $\sim0.05\arcsec$ with respect to that returned by the spectroastrometric modeling. However, the spectroastrometric map probes a spatial region of radius $\sim 0.2\arcsec$, whereas the rotation curves probe distances up to $\sim3\arcsec$  from the disk center; therefore, we expect the former to give a more robust indication of the center position than the latter.

A further constraint that the spectroastrometry can place on the rotation curves analysis is the position of the centers of the three slits on the plane of the sky. As previously described,  when we combine the three spectroastrometric curves to obtain the spectroastrometric map, we  obtain an estimate of the positions of the slit center with an accuracy better than that of the telescope pointing (see G10 and G11 for details).
 
Therefore, we repeated the rotation curves modeling imposing geometrical constraints obtained from the spectroastrometric analysis, namely the slit center positions and the values of $\theta_{LON}$ and $(x_c,y_c)$  (``Fit-R 1'' in Table \ref{tab1}). In fact, this produces a higher value of the reduced $\chi^2$ than the original fit (``Fit-R 0'' in Table \ref{tab1}). 
However, the best-fit \Mdark\ and $M/L$ values are the same, within the uncertainties. The differences relative to the ``Fit-R 0'' case ($\sim-0.03$~dex and of $+0.03$~dex, for \Mdark\ and $M/L$, respectively; table \ref{tab1}) can be considered an indication of the influence of the geometrical disk parameters on the derived \Mdark\ and $M/L$ values.

The slightly worse fit quality of ``Fit-R 1'' may also be an indication that that the inner disk probed by the spectroastrometric analysis has a different geometry (in this particular case the disk geometrical center position) than the outer disk, which influences the rotation curves. The rotation curves, in fact, probe the average disk geometric parameters over a spatial region that extends over the central $\sim100\,pc$ whereas spectroastrometry probes the inner $\sim10\,pc$.
 
Regarding the principal parameter of the modeling, \Mdark, we note that the rotation curve fits give a  mass estimate larger by $\approx 0.35$~dex with respect to the values returned by the spectroastrometric modeling. This difference is larger than the statistical or systematic uncertainties, but the good agreement between the two modeling approaches with respect to the geometrical disk parameters gives confidence that the results of the fitting are robust. We therefore attribute the difference in  \Mdark\ to the different spatial scales probed by the two methods. For the rotation curves approach, as previously noted, the spatial resolution of the observations sets the minimum radius at which the gas rotation curve can be probed. Therefore, with this approach, we can only measure the total dynamical mass enclosed within a radius equal to half the spatial resolution, in this case, corresponding to $\sim0.35\arcsec$ or $7$~pc. On the other hand, the spectroastrometric approach is able to probe the gas rotation curve at smaller radii, allowing us to measure the total dynamical mass enclosed  within ${\sim0.1\arcsec}$ ($\sim2$~pc, see Fig.~\ref{figfit}). We conclude that the difference in the  mass measured by the two approaches ($7.6\times10^6M_{\odot}$ from spectroastrometry and $1.7\times10^7M_{\odot}$ from rotation curves) is most likely  due to an extended mass distribution within the inner $\sim0.35\arcsec$ ($7$~pc), which  is large compared to the BH mass ($\sim1.7\times10^6M_{\odot}$).
 %: fig masses
  \begin{figure}[!t]
  \centering
  \includegraphics[width=\linewidth]{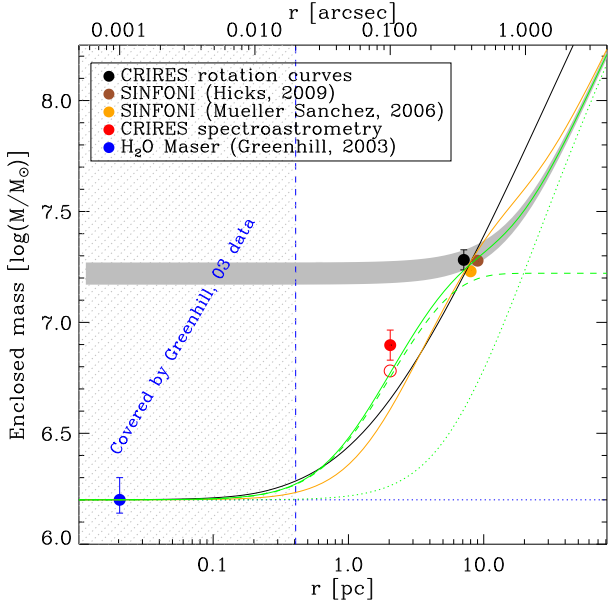}
     \caption{Dynamical mass measurements in the nuclear region of the Circinus galaxy. Measurements are marked by filled circles of different colors and are identified in the legend. The red filled circle represents the CRIRES spectroastrometry mass measurement obtained assuming a disk with $i=61^\circ$  with respect to the line of sight, same as   large scale measurements; the open red circle represents the spectroastrometry mass measurement obtained assuming an edge-on disk, same as the maser disk. 
     The region shaded by gray dotted lines is characterized by $r<20$\,mas and is the region probed by the $\mathrm{H_2O}$ maser rotation curve, where the dynamical mass is mostly unresolved and belongs to the BH. The enclosed mass within this region must be dominated by the BH, since no evidence for extended mass is seen in the maser rotation curve.
The gray stripe represents the radial total mass distribution of the best-fit model for the CRIRES rotation curves (taking into account errors on \Mdark\ and $M/L$). 
The  orange line represent the total mass distribution which is a combination of BH mass,  {nuclear} (4pc size) and extended mass distribution, with the  mass distribution derived by  \cite{Muller-Sanchez:2006}.  {The nuclear mass distribution corresponds to what we measure as} \Mdark.
The  green line is similar to the orange line but the  {nuclear} mass distribution has now a size of 2pc.
The dashed green line is the contribution of BH mass and nuclear mass distribution, while the dotted line represents BH mass and extended mass distribution.
The continuous black line represents the radial mass distribution of the nuclear star cluster at the center of the Milky Way \citep{Genzel:2010a} combined with the BH mass \citep{Greenhill:2003a} and rescaled to match the measurements at $\sim 10$\, pc scale.  }
        \label{figmass}
  \end{figure}

\section{The nuclear mass distribution} 

The most important result from our attempts to measure the mass of the BH in Circinus is that the \Mdark\ value estimated from rotation curve modeling is $\sim 0.35$~dex higher than the value estimated from spectroastrometric modeling. Moreover, the BH mass measured by \citealt{Greenhill:2003a} from the $\mathrm{H_2O}$ maser spectroscopy ($1.7\times10^6M_{\odot}$)  is even lower ($\sim0.7-1$~dex) than our estimates.
The analyses based on rotation curves and spectroastrometry can robustly probe gas rotation and thus the mass distribution, at different distances from the dynamical center. The lower limits on these distances (``limiting inner radii'') are set by the instrumental setup. In the case of rotation curves the limiting inner radius is half the instrumental spatial resolution and corresponds to $\sim0.35\arcsec$ or  {${7}$\,pc} for our Circinus data. In the case of the spectroastrometric analysis the limiting inner radius is a fraction of the spatial resolution and corresponds to $\sim0.1\arcsec$ or $2$\, pc (right panel of Fig.~\ref{figfit}). Therefore the different measurements of the  {unresolved dynamical mass} indicate the existence of an extended mass distribution, revealed by the different spatial scales probed.
 {This extended mass distribution could not be identified in the K-band continuum light profile, also for the presence of the strong hot dust emission within the inner 0\farcs2 \citep{Prieto:2004a}. It was only} revealed as  {unresolved mass} in both our dynamical analyses.

In Fig.~\ref{figmass} we summarize existing dynamical mass measurements performed in the nuclear region of the Circinus galaxy. The enclosed mass is plotted as a function of distance from the dynamical center, which is assumed to coincide with the BH location. The blue point represents the BH mass measurement by \cite{Greenhill:2003a} from modeling of the $\mathrm{H_2O}$ maser rotation curve and is located at the inner radius probed by the maser rotation curves. The maser disk is almost edge-on.
The gray region characterized by $r<20$\,mas is the region probed by the $\mathrm{H_2O}$ maser rotation curve, where the dynamical mass is mostly unresolved and belongs to the BH. The outer limit of this region is set by the maximum radius probed by the maser rotation curve.
The red filled circle represents the CRIRES spectroastrometry mass measurement obtained assuming a disk with $i=61^\circ$  with respect to the line of sight. This is the inclination which is determined from the analysis of the rotation curves at larger scales, and is close to the inclination of the galactic disk.
Since the scales probed by spectroastrometry are intermediate between those of the galactic  and maser disks, we have also considered the possibility that the spectroastrometry is probing rotation from an edge-on disk (open red circle). Clearly, even if the disk were edge-on, the observed kinematics at $\sim 2$pc-scales would still require the presence of extended mass. The black circle represents the mass measurement from the CRIRES rotation curves, and is located at half the spatial resolution. Finally, the brown and orange points represent measurements of dynamical masses enclosed in 8\,pc and 9\,pc radii by  \cite{Muller-Sanchez:2006} and \cite{Hicks:2009}, respectively, using SINFONI AO assisted observations. It is clear that the measurements from SINFONI and CRIRES agree within the errors. 
The gray stripe represents the radial total mass distribution of the best-fit model for the CRIRES rotation curves (taking into account errors on \Mdark\ and $M/L$). This is the mass distribution one would have derived considering only the K band surface brightness profile.

Inspection of Fig.~\ref{figmass} reveals the existence of a spatially extended mass distribution within the inner 10\,pc of the nuclear region of Circinus.
It also shows the capability of spectroastrometry to improve the spatial resolution by a factor $\sim 3$, providing constraints on the extended mass distribution but,
unfortunately, it is still not possible to measure the BH mass. Indeed the stellar mass distribution recovered from the NICMOS K-band continuum light profile does not account for the mass distribution in the inner 10\,pc and leads to an overestimated unresolved  {dynamical} mass. 

The extended mass within 10\,pc might be accounted for by stars not identified in the NICMOS image due to the presence of the hot dust emission or might be accounted for by cold gas. The AO-assisted SINFONI observations by \cite{Muller-Sanchez:2006} provide the surface brightness profiles of hot dust and stellar emission, as well as those of 
Br$\gamma$ and H$_2$. As shown in their figures 3 and 4, the stellar emission is more extended than the hot dust and gaseous components. Moreover, the stellar component if offset by $\sim 0\farcs2$ and its mass is $\sim 10^{6} M_\odot$ making it unlikely that \Mdark\ can be accounted for by stars whose emission is hidden in NICMOS by the strong hot dust emission.
On the other  {hand}, they find that Br$\gamma$ and H$_2$ surface brightness distributions are consistent with an exponential disk with scale radius $r_d \simeq 4$\,pc. Moreover \cite{Muller-Sanchez:2006} find that the nuclear H$_2$ luminosity corresponds to a molecular gas mass of $1.7\times 10^7\,M_\odot$,  fully consistent with the dynamical mass within 10\,pc. Therefore it seems that the extended mass distribution within 10\,pc might be accounted for by a disk of mostly molecular gas.
In order to verify this hypothesis, we plot with an orange line in Fig.~\ref{figmass} the mass distribution obtained by considering the BH mass by \cite{Greenhill:2003a}, the stellar mass distribution from the NICMOS K band surface brightness profile and an exponential disk with $r_d = 4$\,pc and mass $1.7\times 10^7\,M_\odot$. This is consistent with the analysis of \cite{Muller-Sanchez:2006} but fails to reproduce our spectroastrometric measurements because this mass distribution is too extended.
A much better agreement is found considering  an exponential disk with $r_d = 2$\,pc and mass $1.4\times 10^7\,M_\odot$ which is shown by the green solid line. For comparison, the dashed and dotted green lines represent the relative contribution of the gas and stellar masses (each of them added to the BH mass). This is perfectly consistent with the results by \cite{Muller-Sanchez:2006} considering the uncertainties in converting the emission line surface brightness distributions in mass distribution.  

The spectroastrometric measurements probing smaller scales thus indicate that the mass distribution is more concentrated than is inferred from the Br$\gamma$ and H$_2$ surface brightness distributions. 
The mass which is unresolved even by spectroastrometry has a large average density of 1.4$\times 10^5\,\mathrm{M_\odot\,pc^{-3}}$ and is thus expected to host significant amounts of star formation. Indeed  \citep{Muller-Sanchez:2006} found that significant star formation is ongoing within the central 8\,pc and identified the molecular medium with the torus required by the AGN unified model and believed to exist in the nucleus of Circinus, a Seyfert 2 galaxy.  

From the green line it is also possible to infer that the contribution of gas and stars to the mass distribution equals that of the BH mass at $\simeq 1$\,pc which is then to be considered as the radius of the gravitational sphere of influence of the BH. This is perfectly consistent with the independent estimate based on the stellar velocity dispersion.
For a stellar velocity dispersion of $\sigma\simeq 75$ km/s (e.g., \citealt{graham:2008a}) the sphere of influence is $r_{BH} = G M_{BH}/\sigma^2 = 1.2 $\,pc remarkably similar to what we have just found.

Finally, the black line represents the mass profile which is obtained by considering the BH mass and the star cluster in the Milky way \citep{Genzel:2010a} with a mass rescaled to match the observed enclosed mass at 10\,pc. 
\cite{Genzel:2010a} model the Milky way nuclear star cluster with a double power law density distribution characterized by a break radius of $0.25pc$, inner power law $r^{-1.3}$ , outer power law $r^{-1.8}$ and density at the break radius $1.35\times10^6M_{\odot}pc^{-3}$.

The comparison with the Milky Way is suggested by the similarity of morphological types between the two galaxies and indicates that, overall, the nuclear mass distributions at $r<10$\,pc scales are not significantly different. It is intriguing to think that, if all the molecular gas in Circinus is converted into stars then it will produce a nuclear star cluster similar to that observed in the Milky way and  {also commonly} in late type spirals.
Since the molecular material is likely associated with the torus invoked by the AGN unified model, it is also intriguing to speculate that there exists an evolutionary track in which dense nuclear molecular tori are converted into nuclear star clusters. 

\section{Conclusions}
We have presented new CRIRES spectroscopic observations of the Br$\gamma$ emission line in the nuclear region of the Circinus galaxy, obtained with the aim of measuring the BH mass using the spectroastrometric technique. The Circinus galaxy is an ideal benchmark for the spectroastrometric technique, given its proximity and the existence of a secure BH measurement obtained from observations of the nuclear H$_2$O maser disk.

The kinematical data have been analyzed using both the classical method based on the rotation curves and with a new method developed by us and based on spectroastrometry.

The classical method indicates that the gas disk has  inclination and position angles consistent with those of the large scale galactic disk. The disk rotates in a gravitational potential resulting from an extended stellar mass distribution and a spatially unresolved  {dynamical} mass of $(1.7\pm 0.2)\times 10^7\,M_\odot$, {concentrated within $r<7$ pc}, {corresponding to the seeing-limited resolution of the observations}. The estimates of inclination, position angle and mass are consistent with previous measurements based on SINFONI spectroscopy.

The spectroastrometric method is capable of probing the gas rotation at scales which are a factor $\sim 3.5$ smaller than those set by the spatial resolution. The  {dynamical} mass that is spatially unresolved at  the scales probed by spectroastrometry is a factor $\sim 2$ smaller, $7.9^{+1.4}_{-1.1}\times 10^6\,M_\odot$, than that recovered from the rotation curve analysis, indicating that spectroastrometry has spatially resolved the nuclear mass {distribution down to 2-pc scales}. This unresolved mass is still a factor $\sim 4.5$ larger than the BH mass measurement obtained with the H$_2$O maser emission indicating that, even with spectroastrometry, it has not been possible to resolve the sphere of influence of the BH.

Based on the detailed analysis by \cite{Muller-Sanchez:2006}, this dark mass distribution is likely composed of warm molecular gas as revealed by the  {H$_2$ 1-0 S(1) emission at 2.12$\,\mu$m}. This medium has a large density, is clumpy and forming stars, and has been tentatively identified with the molecular torus required by the AGN unified model, which prevents a direct view of the central BH in Circinus.
The mass distribution is similar in shape to that of the nuclear star cluster of the Milky Way. We tentatively speculate that molecular tori, forming stars at a high rate, might be the precursors of the nuclear star clusters that are common in late type spirals.

\begin{acknowledgements}
We wish to honor the memory of our great friend and colleague David Axon. He will be greatly missed by all of us.
We are grateful to Roberto Maiolino for providing us the reduced HST NICMOS images.
AG and AM acknowledge financial support from the Italian National Institute for Astrophysics by INAF CRAM 1.06.09.10
\end{acknowledgements}

\bibliographystyle{aa} % style aa.bst
%\bibliography{biblio_extn0.1.bib}

\end{document}